\documentclass[%
 reprint,
superscriptaddress,
 amsmath,amssymb,
 aps,
pra,
longbibliography
]{revtex4-2}
\usepackage{xcolor}
\usepackage[normalem]{ulem}
\usepackage{graphicx}
\usepackage{dcolumn}
\usepackage{bm}
\usepackage{hyperref}
\hypersetup{
    colorlinks=true
}

\begin{document}

\preprint{APS/123-QED}

\title{Phonon condensation and cooling
via nonlinear feedback}

\author{Xu Zheng}
\thanks{Current address: School of Physical and Mathematical Sciences, Nanyang Technological University, Singapore 637371, Singapore}
\email{Xu.Zheng@Colorado.Edu}
\affiliation{Department of Physics, University of Colorado, Boulder, CO, 80309, USA}
\author{Baowen Li}
\thanks{Current address: Department of Physics, Southern University of Science and Technology, Shenzhen 518055, China}
\email{Baowen.Li@Colorado.Edu}
\affiliation{Paul M. Rady Department of Mechanical Engineering, University of Colorado, Boulder, CO, 80309, USA}
\affiliation{Department of Physics, University of Colorado, Boulder, CO, 80309, USA}
\date{\today}

\begin{abstract}
We propose a method to control the energy distribution in multimode mechanical systems using a single nonlinear feedback loop. We demonstrate that this feedback mechanism simultaneously amplifies the fundamental vibrational mode while suppressing all higher-order modes, effectively channeling energy into the lowest-frequency mode. This process mimics the energy redistribution of Fr\"{o}hlich condensation but is achieved here through a designed feedback force that combines a ``low-pass gain'' and a ``high-pass loss''. In the feedback-induced steady state, the fundamental mode exhibits a phase-space distribution similar to that of a phonon laser, characterized by a ring shape and amplitude squeezing. Additionally, we show that the linewidth of the fundamental mode is narrowed by an order of magnitude, corresponding to a significant enhancement in phase coherence. This scheme offers a robust approach to generating coherent mechanical states and phonon lasing without the need for optical gain media or intrinsic material nonlinearities.
\end{abstract}

\maketitle


\emph{Introduction}. -- 
The manipulation and control of phonons and vibrational energy are of significant interest for both engineering applications and fundamental research. On the one hand, coherent phonons show great potential in applications ranging from conventional nondestructive testing \cite{dekorsy2000coherent,ruello2015physical}, high-resolution imaging and sensing \cite{poyser2015coherent}, to quantum information processing \cite{ruskov2012coherent,gustafsson2014propagating,bienfait2019phonon}. On the other hand, controlling incoherent phonons is crucial for noise reduction \cite{liu2020review}, thermoelectric energy conversion \cite{takabatake2014phonon}, and thermal management \cite{li2012colloquium,li2021transforming}.

In the context of micromechanical and nanomechanical resonators, which are widely used for ultrasensitive sensing \cite{rugar2004single,yang2006zeptogram,burg2007weighing}, acoustic actuation \cite{masmanidis2007multifunctional,feng2008self,wen2020coherent}, information processing \cite{mahboob2008bit,unterreithmeier2009universal,tadokoro2018driven}, and biological imaging \cite{tamayo2001high,shekhawat2005nanoscale,tetard2008imaging}, amplifying vibration amplitude and narrowing phonon linewidth are critical for performance. Active linear feedback control, where the feedback force is proportional to the measured mechanical displacement or velocity with a specific phase difference, is a well-known technique for achieving these two goals \cite{poggio2007feedback,ohta2017feedback}. Depending on the phase difference, either positive or negative feedback can be realized. This method relies on real-time monitoring of mechanical motion and is highly effective when the resonator operates as a single-mode system. 

However, mechanical resonators inherently possess a spectrum of normal modes. A standard linear feedback loop typically results in the simultaneous amplification or damping of multiple modes \cite{ohta2017feedback,sommer2019partial}. For applications such as energy harvesting and phonon lasing, where energy concentration in a single selected mode is desired, a fundamental question arises: is it possible to amplify a specific mode while simultaneously cooling all others using a single feedback loop?

This selective amplification of the fundamental mode accompanied by the suppression of higher-order modes is closely related to the phenomenon of Fr\"{o}hlich condensation \cite{frohlich1968bose,frohlich1968long,frohlich1970long,wu1977bose,wu1978cooperative,wu1981frohlich,reimers2009weak,preto2017semi,zhang2019quantum,zheng2021froh}. In this process, vibrational energy in a collection of oscillators condenses into the fundamental mode once the external energy supply exceeds a critical threshold. The essential mechanism driving Fr\"{o}hlich condensation is energy redistribution induced by nonlinear couplings, which is predicted to exist in some biological and optomechanical systems but has not yet been experimentally verified.

In this paper, we propose a method to realize a similar condensation phenomenon in multimode mechanical systems using a single nonlinear feedback loop, bypassing the need for intrinsic nonlinearities required in conventional Fr\"{o}hlich condensation. We demonstrate that this feedback mechanism not only channels energy into the fundamental mode but also induces a steady state characterized by strong amplitude coherence, evidenced by a ring-shaped phase space distribution. Furthermore, the phase coherence of the fundamental mode is significantly enhanced. 

\begin{figure}[t]
    \centering
    \includegraphics[width=0.7\linewidth]{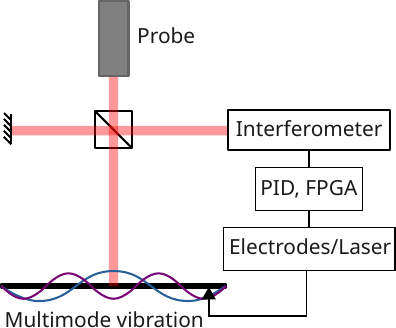}
    \caption{Sketch of the multimode system considered. The reflected optical field provides information on the collective displacement of the resonator. Based on the detected signal, the feedback loop determines the drive applied to the resonator. The feedback force can be realized using optomechanical, photothermal, or electromechanical effects.}
    \label{fig:config}
\end{figure}

\emph{Model}. -- We consider a multimode mechanical resonator. The system configuration is illustrated in Fig.~\ref{fig:config}. In a typical experimental setup (optomechanical or electromechanical), the mechanical displacement is monitored via a probe laser. The resulting signal is processed in real-time by a field-programmable gate array (FPGA)-based digital controller to compute the nonlinear feedback function \cite{wang2023fastfeedback}. The feedback force is then applied to the resonator by modulating the intensity of a drive laser (via radiation pressure or photothermal stress) \cite{ohta2017feedback,Guo2019feedback,guo2023active} or by applying a voltage to electrodes (electromechanical force) \cite{poggio2007feedback}.


The dynamics of the resonator are governed by the classical Langevin equations of motion:
\begin{align}
    \dot{Q}_j&=\omega_jP_j,\nonumber\\
    \dot{P}_j&=-\omega_jQ_j-\gamma_jP_j+\xi_j+H_{\text{fb}}^{(j)},
    \label{eq:eoms}
\end{align}
where $\omega_j$ is the angular frequency, and $Q_j=\sqrt{\frac{m}{k_BT}}\omega_jq_j$ and $P_j=\frac{p_j}{\sqrt{mk_BT}}$ denote the dimensionless displacement and momentum of the $j$th normal mode, normalized such that the thermal equilibrium energy $\frac{1}{2}(Q_j^2+P_j^2)$ is unity. Here, $\gamma_j$ represents the damping rate, $\xi_j$ is the thermal noise force, and $H_{\text{fb}}^{(j)}$ is the feedback force acting on the $j$th mode. In the high-temperature limit ($k_BT\gg\hbar\omega_j$), the thermal noise satisfies the fluctuation-dissipation relation $\langle\xi_j(t)\xi_j(t^{\prime})\rangle=2\gamma_j\delta(t-t^{\prime})$. The feedback force $H_{\text{fb}}^{(j)}$ depends on the measured collective displacement $Q=\sum_jQ_j$. To achieve phonon condensation in the fundamental mode, we design the following nonlinear feedback loop:
\begin{align}
    F_I&=\int_{0}^tQ(s)ds,\nonumber\\
    F_D&=\dot{Q},\nonumber\\
    H_{\text{fb}}^{(j)}&=-g_j\tanh{[\omega_{\text{fb}}(F_I^2F_D+3Q^2F_I)]},
    \label{eq:feedbackloop}
\end{align}
where $g_j$ is the feedback gain and $\omega_{\text{fb}}$ is a reference frequency ensuring dimensional consistency. The terms $F_I$ and $F_D$ represent the integral and derivative components, respectively, which can be obtained from a proportional–integral–derivative (PID) controller. A hyperbolic tangent function is incorporated to saturate the feedback force within the range $\pm g_j$. Physically, this feedback loop introduces two competing effects: a term proportional to $F_D\sim\omega_j$ acting as a ``high-pass loss'', meaning high-frequency modes experience stronger damping, and a term proportional to $F_I\sim 1/\omega_j$ acting as a ``low-pass gain'', meaning low-frequency modes experience stronger amplification. The interplay of these effects results in the simultaneous cooling of high-frequency modes and amplification of the fundamental mode, mimicking the energy redistribution process characteristic of Fr\"{o}hlich condensation. 

To understand the mechanism of the feedback loop, we first analyze a simplified form $H_{\text{fb},0}^{(j)}$ where the hyperbolic tangent is replaced by the identity function, i.e., $H_{\text{fb},0}^{(j)}=-g_j\omega_{\text{fb}}(F_I^2F_D+3Q^2F_I).$ This approximation is valid in the weak feedback regime. By introducing slowly varying amplitude and phase variables via the ansatz
\begin{align}
    Q_j(t)&=a_j(t)\cos{(\omega_jt+\chi_j(t))}
    \label{eq:complexamp}
\end{align}
with $a_j(t)$ [$\chi_j(t)$] being slowly varying amplitudes (phases) ($\dot{a}_j\ll\omega_ja_j,$ $\dot{\chi}_j\ll\omega_j$), we can simplify the amplitude equations as
\begin{align}
    \dot{a}_{j}=-\frac{\gamma_j}{2}a_j+\sum_i\frac{g_j\omega_{\text{fb}}}{4\omega_i^2\omega_j}(\omega_i^2-\omega_j^2)|a_i|^2a_j+\Xi_j.
    \label{eq:amplitudeequations}
\end{align}
In the derivation, we have assumed a high-quality factor $\omega_j\gg\gamma_j$, ignored off-resonant terms, and averaged the thermal noise $\xi_j(t)$ over the fast dynamics,
\begin{align}
    \Xi_j(t)=\frac{\omega_j}{2\pi}\int_{t-\pi/\omega_j}^{t+\pi/\omega_j}ds\xi_j(s)e^{i\omega_js}.
\end{align}
The slowly varying noise $\Xi_j(t)$ satisfies
\begin{align}
    \langle\Xi_j(t)\Xi^{\ast}_j(t^{\prime})\rangle=2\gamma_j\delta(t-t^{\prime}).
    \label{eq:correlation}
\end{align}
The detailed derivation is provided in Appendix \ref{appendix:amplitudeequations}.

\begin{figure*}[t]
    \centering
    \includegraphics[width=0.95\linewidth]{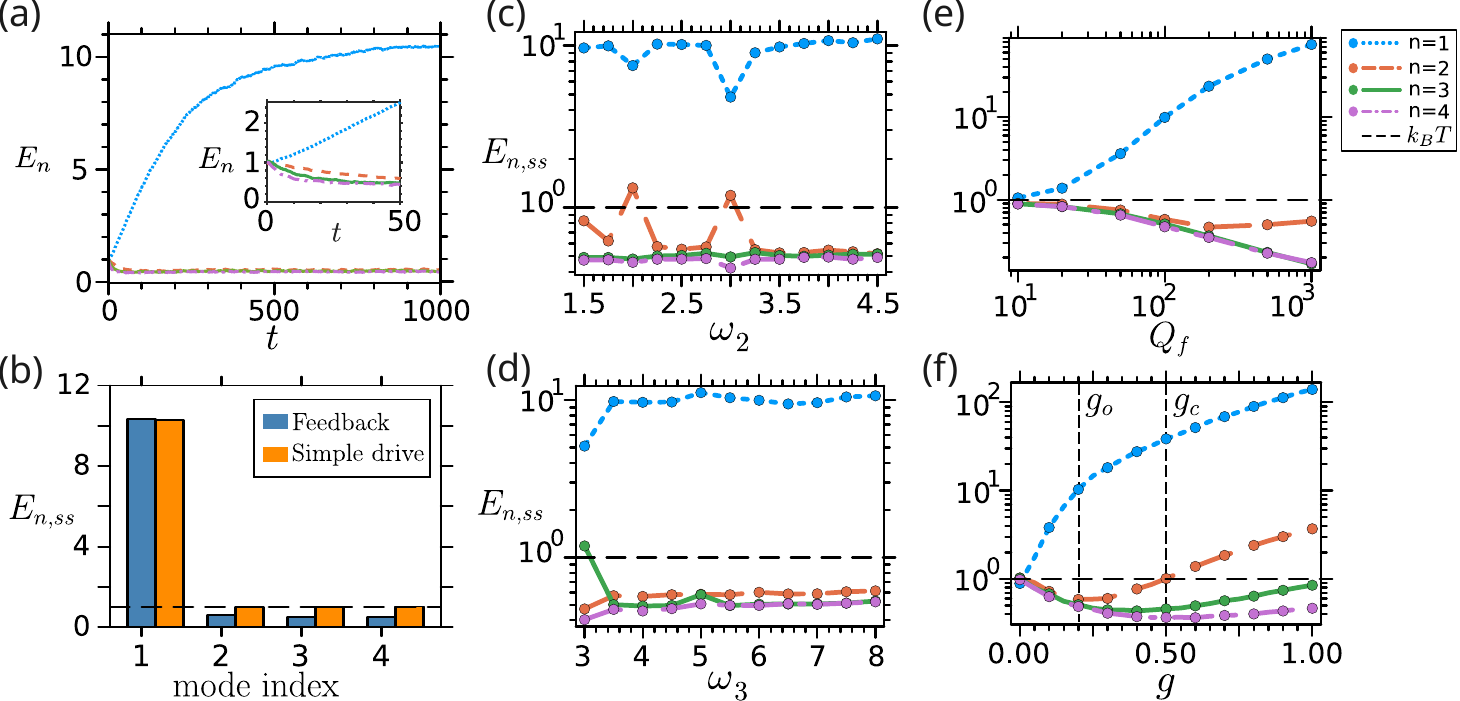}
    \caption{(color online). Phonon condensation in a system with $N=4$ modes. (a) Vibration energy as a function of time. The dimensionless vibration energy at thermal equilibrium ($t=0$) is one. The inset shows a zoomed-in view. The lowest mode is amplified by a factor of 10, while the other three modes are cooled to 0.59, 0.51, and 0.50, respectively. (b) Steady-state mode energies under nonlinear feedback (blue) and under a simple resonant drive at $\omega_1$ (orange). The drive amplitude is chosen so that the mean energy of the fundamental mode is the same in both cases. (c) and (d) Steady-state energy of each mode as a function of the second and third mode frequencies $\omega_2$ and $\omega_3$, respectively. (e) Dependence of steady-state energy on the mechanical quality factor $Q_f$. (f) Dependence of steady-state energy on the feedback gain $g$. In our simulation, the frequencies of each mode are chosen according to the continuum elasticity theory $\omega_j=kc_j$, where the values $c_j$ are obtained by solving the equation $\cos{\sqrt{c_j}}\cosh{\sqrt{c_j}}=1$ and $k$ is a constant depending on the geometry and material of the resonator \cite{landau1986course}. The first four modes satisfy $\omega_2/\omega_1=2.75$, $\omega_3/\omega_1=5.13$, $\omega_4/\omega_1=8.75$. The other parameters are $\gamma_j/\omega_j=10^{-2}$, $g_j=g=0.2$, $\omega_{\text{fb}}/\omega_1=1$.}
    \label{fig:energy}
\end{figure*}

\emph{Phonon condensation in the fundamental mode}. -- From the amplitude equations (\ref{eq:amplitudeequations}), we can define the effective damping rate 
\begin{align}
    \tilde{\gamma}_j=\gamma_j+\sum_i\frac{g_j\omega_{\text{fb}}}{2\omega_i^2\omega_j}(\omega_j^2-\omega_i^2)|a_i|^2,
    \label{eq:effectiverate}
\end{align}
where the second term is induced by the nonlinear feedback. To understand the system dynamics, we first consider the case of $N=2$ modes. For $N=2$, the feedback term is negative for the lower-frequency mode ($\omega_1 < \omega_2$), reducing its effective damping ($\tilde{\gamma}_1 < \gamma_1$). In contrast, it is positive for the higher-frequency mode, enhancing its damping ($\tilde{\gamma}_2 > \gamma_2$). Using the approximate steady-state energy relation $E_{j,ss} \approx \gamma_j/\tilde{\gamma}_j$, we predict amplification for the first mode ($E_{1,ss} > 1$) and cooling for the second ($E_{2,ss} < 1$) relative to thermal equilibrium. Generalizing to $N>2$ modes, the fundamental mode always experiences reduced damping ($\tilde{\gamma}_1 < \gamma_1$) since $\omega_1^2 - \omega_i^2 < 0$ for all $i > 1$. While a full analytical solution for all modes is complex, a self-consistent argument can be made by assuming phonon condensation occurs, i.e., the energy of the fundamental mode far exceeds that of others. In this regime, the feedback interaction is dominated by the term proportional to $(\omega_j^2-\omega_1^2)|a_1|^2$. Consequently, the $N$-mode system effectively decouples into a set of pairwise interactions between the fundamental mode and each higher mode. Applying the $N=2$ analysis to these pairs yields $E_{1,ss} > 1 > E_{j,ss}$ for all $j > 1$, validating the initial assumption.

Although derived for the simplified feedback $H_{\text{fb},0}^{(j)}$, these conclusions hold for the full feedback $H_{\text{fb}}^{(j)}$ incorporating the saturation function. The hyperbolic tangent function limits the force magnitude without altering its sign, preserving the direction of energy flow. We verify these predictions through numerical integration of the full stochastic equations of motion [Eq. (\ref{eq:eoms})] using the \texttt{DifferentialEquations.jl} package \cite{rackauckas2017differentialequations}. Figure~\ref{fig:energy}a displays the energy evolution for a system with $N=4$ modes. As anticipated, the fundamental mode dominates the long-time dynamics, achieving a tenfold amplification. In contrast, the higher modes are simultaneously cooled to approximately half their thermal energy (0.59, 0.51, and 0.50, respectively). For comparison, we also simulate a simple resonant drive at $\omega_1$ without feedback. The drive amplitude is chosen so that the steady-state mean energy of the fundamental mode matches that of the feedback case. Figure~\ref{fig:energy}b shows that, although a simple drive amplifies the fundamental mode, it leaves the higher modes close to their thermal values. The advantage of the nonlinear feedback is therefore not a larger mean energy of the fundamental mode, but the simultaneous amplification of the fundamental mode and cooling of all higher modes.

To investigate the robustness of phonon condensation, we examine its dependence on various system parameters in Figs.~\ref{fig:energy}c--\ref{fig:energy}f. Figures~\ref{fig:energy}c and \ref{fig:energy}d show the steady-state energy of each mode as a function of the second and third mode frequencies, respectively. We observe that phonon condensation is robust against frequency detuning, occurring for a wide range of frequencies, except when the mode frequencies are commensurate (e.g., $\omega_j/\omega_1 \approx 2, 3$). This breakdown occurs because our derivation assumes incommensurate frequencies and treats frequency differences between different modes as non-resonant terms that can be ignored. In Fig.~\ref{fig:energy}e, we explore the effect of the mechanical quality factor $Q_f\equiv\omega_j/\gamma_j$. The condensation persists over a broad range of $Q_f$ factors, with higher $Q_f$ leading to more efficient condensation. However, the effect diminishes at low $Q_f$, disappearing around $Q_f \approx 10$ for the parameters used in our simulations. Finally, Fig.~\ref{fig:energy}f illustrates the dependence on the feedback gain $g$. We identify an optimal gain $g_o$ and a critical gain $g_c$. For $g < g_o$, increasing the gain enhances the amplification of the fundamental mode and the damping of higher modes. In the intermediate regime $g_o < g < g_c$, the fundamental mode continues to be amplified, but the damping of higher modes becomes less effective, although their energy remains below the thermal energy level. When $g > g_c$, the higher modes begin to heat up, exceeding their thermal energy. This heating behavior at large gains is attributed to the breakdown of the rotating wave approximation used in our theoretical derivation.

\begin{figure}[t]
    \centering
    \includegraphics[width=\linewidth]{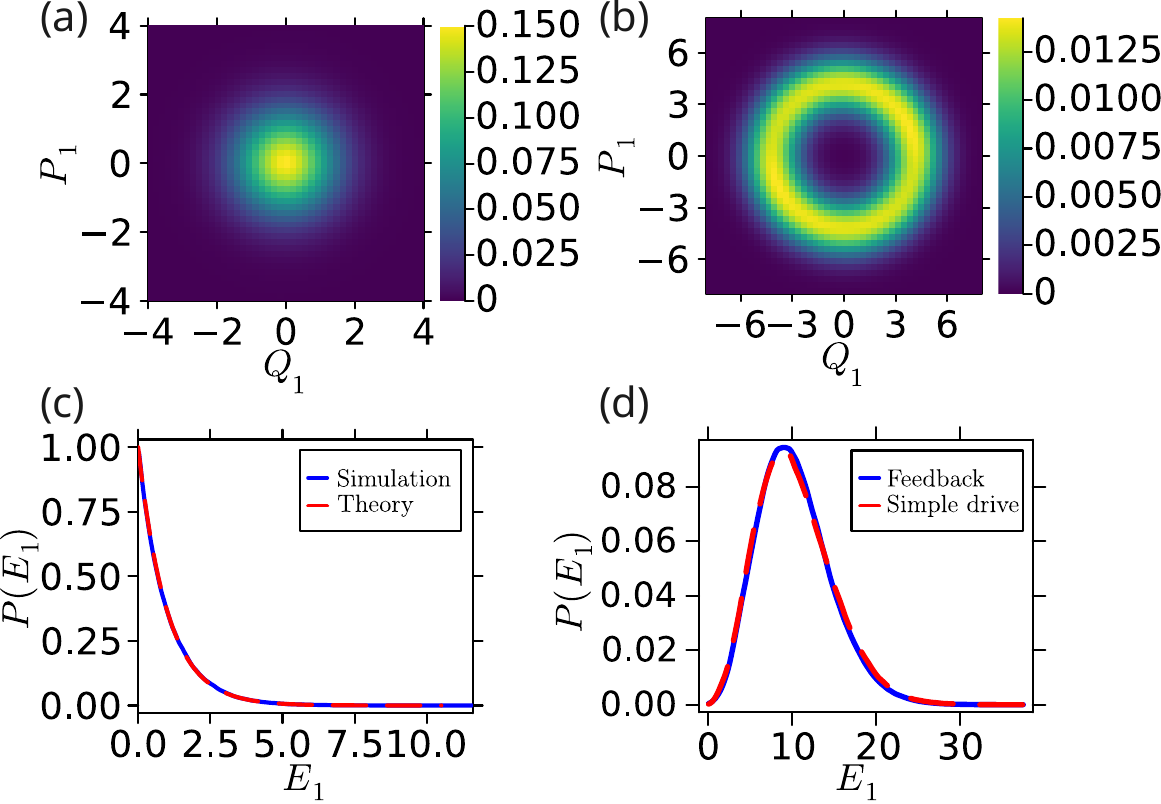}
    \caption{(color online). Statistical properties of the fundamental mode. (a) Phase space distribution of the fundamental mode without feedback. (b) Phase space distribution with feedback. (c) Energy distribution without feedback. (d) Simulated energy distribution of the fundamental mode with feedback (blue solid line) and with a simple drive (red dashed line). The parameters used are the same as those in Fig.~\ref{fig:energy}.
    The variance ($\approx 19$) in (d) is much smaller than the variance ($\approx 110$) of a thermal state with the same mean energy ($\approx 10$), indicating strong amplitude squeezing.
    }
    \label{fig:phononstatistics}
\end{figure}

\emph{Phase space distribution and coherence of the fundamental mode}. -- So far, we have only discussed the vibration energy in each mode and shown that the feedback can give rise to phonon condensation in the fundamental mode. To gain more information about the feedback-induced steady state, we investigate the phase space distribution of the fundamental mode. The phase space distribution is often described by the Wigner function. In the high-temperature limit considered here, the quantum Wigner function is well approximated by the classical probability distribution in phase space, which can be simulated using classical Langevin equations. Figures~\ref{fig:phononstatistics}a and \ref{fig:phononstatistics}b display the phase space distribution of the fundamental mode without and with feedback. Without feedback, the distribution is centered at the origin, consistent with thermal Brownian motion. With feedback, however, the distribution exhibits a ring shape, which is similar to that of a phonon laser \cite{pettit2019optical,wen2020coherent} and indicates the existence of amplitude coherence.

To further compare the statistical properties without and with feedback, we show the energy distribution of the fundamental mode in Figures~\ref{fig:phononstatistics}c and \ref{fig:phononstatistics}d (blue solid line). Without feedback, the energy distribution follows the exponential distribution of thermal Boltzmann statistics. With feedback, the energy distribution shifts from Boltzmann statistics to a distribution whose most probable energy is nonzero. The variance ($\approx 19$) observed in Fig.~\ref{fig:phononstatistics}d exceeds the mean ($\approx 10$), yet remains significantly below the variance ($\approx 110$) of a thermal state with the same mean energy, indicating substantial suppression of thermal amplitude fluctuations. Furthermore, the second-order correlation function $g^{(2)}(0)$ obtained from the energy distribution is $g^{(2)}(0)=\langle E_1^2\rangle/\langle E_1\rangle^2=1.17$, which is close to the $g^{(2)}(0)$ value of the ideal coherent state ($g^{(2)}(0)=1$). 

We also compare this distribution with that produced by a simple resonant drive at $\omega_1$, with the drive amplitude chosen so that the mean energy of the fundamental mode is the same in both cases. For a simple drive, the energy follows a noncentral $\chi^2$ distribution; Appendix~\ref{appendix:simpledrive} gives the analytical derivation and the corresponding energy variance. As shown in Fig.~\ref{fig:phononstatistics}d, the energy distributions for the simple drive (red dashed line) and the feedback case (blue solid line) nearly overlap. This indicates that the feedback protocol preserves the same level of amplitude coherence of the amplified fundamental mode as in the simple-drive case. Its advantage instead lies in combining this near-coherent amplification with simultaneous cooling of all higher modes.

By examining the phase space distribution, we demonstrate the amplitude coherence of the fundamental mode in the feedback-induced steady state. We are also interested in the phase coherence, which can be determined by the linewidth of the noise power spectral density $S_{Q_1Q_1}(\omega)$. The spectral density is obtained by the Fourier transform of autocorrelation function, i.e.,
\begin{align}
    S_{Q_1Q_1}(\omega)=\int_{-\infty}^{+\infty}d\tau\langle \overline{Q_1(t)Q_1(t+\tau)}\rangle e^{i\omega\tau},
    \label{eq:sqq}
\end{align}
where the overline denotes time average over $t$ and the angle brackets denote ensemble average. A numerical method to calculate the spectral density is provided in Appendix \ref{appendix:spectraldensity}. Fig.~\ref{fig:correlation}a shows the spectral density $S_{Q_1Q_1}(\omega)$ without and with feedback. The feedback amplifies the peak value by two orders of magnitude. To demonstrate the enhancement of coherence, we plot the rescaled spectral density in Fig.~\ref{fig:correlation}b. Without feedback, the intrinsic relative linewidth is $\gamma_1/\omega_1=1\times10^{-2}$, corresponding to a quality factor $Q_f = \omega_1/\gamma_1 = 100$. With feedback, the relative linewidth narrows to $\tilde{\gamma}_1/\omega_1=7\times10^{-4}$, yielding an effective quality factor $Q_{\text{eff}} \approx 1428$. This represents an order of magnitude improvement in the coherence of the system.

\begin{figure}[t]
    \centering
    \includegraphics[width=\linewidth]{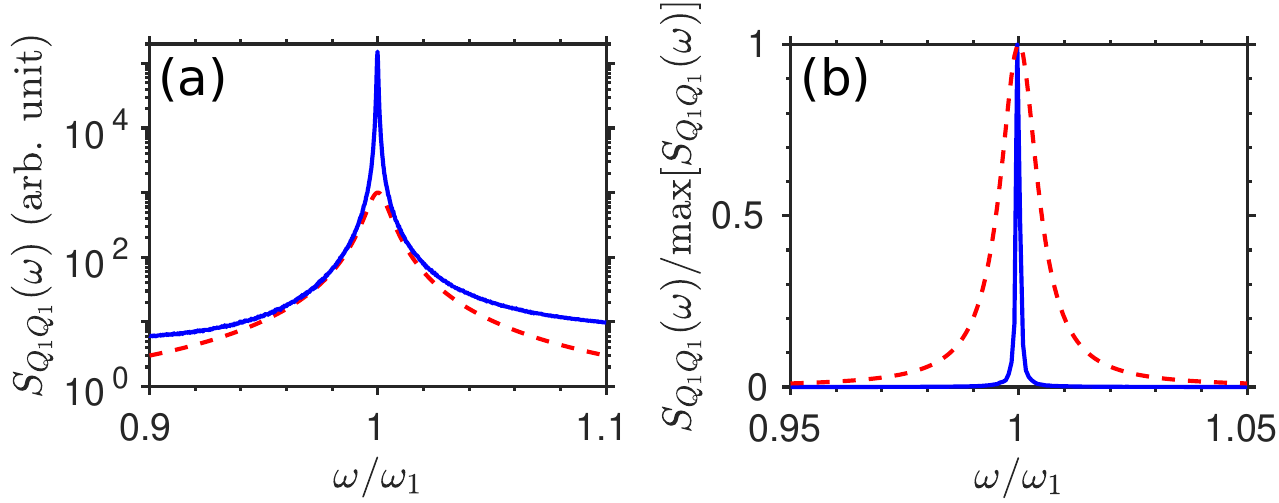}
        \caption{(color online). Noise power spectral density of the fundamental mode. (a) Spectral density on a logarithmic scale. The feedback amplifies the peak value by a factor of 150. (b) Rescaled spectral density. The blue solid lines are the spectral density with feedback, and the red dashed lines are the spectral density without feedback. With feedback, the linewidth is $\tilde{\gamma}_1/\omega_1=7\times10^{-4}$. Without feedback, the intrinsic linewidth is $\gamma_1/\omega_1=1\times10^{-2}$. The parameters used are the same as those in Fig.~\ref{fig:energy}.}
    \label{fig:correlation}
\end{figure}

    \emph{Discussions and conclusions}. -- The energy evolution depicted in Fig.~\ref{fig:energy} closely parallels the phonon number dynamics observed in Fr\"{o}hlich condensation \cite{preto2017semi,zheng2021froh}, illustrating the close connection between our model and Fr\"{o}hlich condensation. This connection can also be seen from the similarity between the amplitude equations (\ref{eq:amplitudeequations}) and the rate equations of phonon numbers in Fr\"{o}hlich's model (see Appendix \ref{appendix:comparison}). In Fr\"{o}hlich's model, there are third-order terms in the Hamiltonian that couple the environment or auxiliary optical field with pairs of vibration modes \cite{wu1977bose,wu1978cooperative,wu1981frohlich,zheng2021froh}, inducing energy redistribution among these modes. In our model, these interaction terms are replaced by the nonlinear feedback loop, where a nonlinear functional of the collective motion $Q=\sum_jQ_j$ induces the interactions between different vibration modes. Thus, our approach offers a pathway to realize Fr\"{o}hlich-like phonon condensation in intrinsically linear mechanical systems. 

To check the experimental feasibility, we use real experimental parameters of an optomechanical system to estimate the maximum pump power required for our feedback scheme to achieve phonon condensation. Considering a GaAs membrane resonator with parameters studied in Ref. \cite{ohta2017feedback}, the first four mechanical modes have frequencies $\Omega_1=160.5$ kHz, $\Omega_2=180.5$ kHz, $\Omega_3=205$ kHz, and $\Omega_4=231$ kHz, respectively. The effective mass is $m_{\text{eff}}\approx10$ ng, and the mechanical quality factor is $Q_f=2000$. At room temperature, the feedback gain $g$ in their experiment is approximately related to the laser power via $g/P \sim 1$ Hz/$\mu$W. From simulations, we find that $g=0.1\Omega_1=16$ kHz is sufficient to concentrate over 80\% of the energy in the fundamental mode. This corresponds to a required laser power of around $16$ mW, which is achievable in current experiments \cite{dong2019more}.

In summary, we have analyzed the prospects for using a nonlinear feedback loop to realize the condensation of phonon or vibration energy in multimode mechanical systems. We have shown that the proposed feedback decreases the effective damping rate of the fundamental mode while increasing the effective damping rate of other modes, resulting in the amplification of the fundamental mode and the damping of all others.  
For the statistical properties and coherence of the fundamental mode, the ring-shaped phase space distribution and significantly narrowed linewidth reveal intriguing similarities between the feedback-induced state and a phonon laser. We note that squeezed states in mechanical resonators have been investigated for decades, for instance, in the early theoretical work by Hu and Nori \cite{hu1996squeezed}. Distinct from these single-mode studies, our work demonstrates that amplitude squeezing can be realized in a multimode system via nonlinear feedback. The key advantages of our scheme include the simultaneous damping of all higher-order modes, ensuring that the energy is effectively concentrated into the coherent fundamental mode, and the elimination of the need for intrinsic nonlinearities. 
These features suggest that the nonlinear feedback loop could be used for sensitive sensing and for the design of novel monochromatic phonon lasers that do not require two-level gain media.

While we have used continuum elasticity theory to model the normal modes of the mechanical resonator, the proposed feedback loop is applicable to general mechanical systems with incommensurate frequencies. Potential platforms include nanoelectromechanical systems (NEMS), levitated nanoparticles in optical tweezers, collective motions of cold atoms or ions in potential traps, and photonic crystals. Further development of the current model could replace the harmonic oscillators with more realistic nonlinear oscillators or self-sustained oscillators (e.g., the Van der Pol oscillator). There are many interesting phenomena in coupled self-sustained oscillators, such as synchronization \cite{pikovsky2003synchronization} and mode competition \cite{jonsson2008self,kemiktarak2014mode,zhang2018mode}. Our proposed feedback could be used to control these phenomena. 

In our system, the essential aspect is the detailed form of the feedback loop, which determines how effectively phonon or energy condensation can be achieved. While the proposed feedback strategy works well, other feedback strategies might provide similar or even better results. Searching for improved feedback strategies, especially with the help of rapidly developing machine-learning methods, is a promising direction \cite{sommer2020prospects}.
In addition, the effects of time delay and phase differences in the feedback loop are interesting topics that deserve further study.

\appendix

\section{Amplitude equations}
\label{appendix:amplitudeequations}
We start from the ansatz:
\begin{align}
    \label{eq:xja}
    Q_j(t)&=a_j(t)\cos{(\omega_jt+\chi_j(t))}\\
    \dot{Q}_j(t)&=-a_j(t)\omega_j\sin{(\omega_jt+\chi_j(t))}
    \label{eq:xjb}
\end{align}
with $a_j(t)$ and $\chi_j(t)$ being slowly varying amplitudes and phases ($\dot{a}_j\ll\omega_ja_j, \dot{\chi}_j \ll\omega_j\chi_j$). To ensure the consistency of Eqs. (\ref{eq:xja}) and (\ref{eq:xjb}), we impose the constraint:
\begin{equation}
    \dot{a}_j\cos{(\omega_jt+\chi_j)}-a_j\dot{\chi}_j\sin{(\omega_jt+\chi_j)}=0,
    \label{eq:thirdequation}
\end{equation}
which is obtained by taking the time derivative of Eq. (\ref{eq:xja}) and equating it to Eq. (\ref{eq:xjb}). Further differentiating Eq. (\ref{eq:xjb}) with respect to $t$, we obtain the equation for $\ddot{Q}_j$ (or $\omega_j\dot{P}_j$):
\begin{align}
    \omega_j\dot{P}_j=&-a_j\omega_j^2\cos{(\omega_jt+\chi_j)}-\dot{a}_j\omega_j\sin{(\omega_jt+\chi_j)}\nonumber\\
    &-a_j\dot{\chi}_j\omega_j\cos{(\omega_jt+\chi_j)}
    \label{eq:secondxj}
\end{align}
Substituting Eqs. (\ref{eq:xja}), (\ref{eq:xjb}), and (\ref{eq:secondxj}) into Eq. (\ref{eq:eoms}), we obtain a nonlinear equation for $a_j$ and $\chi_j$:
\begin{align}
    \dot{a}_j\sin{(\omega_jt+\chi_j)}+a_j\dot{\chi}_j\cos{(\omega_jt+\chi_j)}\nonumber\\
    =-\gamma_ja_j\sin{(\omega_jt+\chi_j)}-H^{\text{fb},0}_j-\xi_j
    \label{eq:eomwithforce}
\end{align}
Combining Eqs. (\ref{eq:thirdequation}) and (\ref{eq:eomwithforce}) yields two first-order differential equations for $a_j$ and $\chi_j$:
\begin{align}
    \dot{a}_j&=-\gamma_ja_j\sin^2{(\omega_jt+\chi_j)}-H^{\text{fb},0}_j\sin{(\omega_jt+\chi_j)}\nonumber\\
    &~~~~-\xi_j\sin{(\omega_jt+\chi_j)}\\
    \dot{\chi}_j&=-\gamma_j\sin{(\omega_jt+\chi_j)}\cos{(\omega_jt+\chi_j)}\nonumber\\
    &~~~~-\frac{H^{\text{fb},0}_j}{a_j}\cos{(\omega_jt+\chi_j)}-\frac{\xi_j}{a_j}\cos{(\omega_jt+\chi_j)}
\end{align}
Averaging these equations over the fast oscillation period (denoted by $\langle\cdots\rangle_T$), we obtain the slow dynamics of the system:
\begin{align}
\label{eq:chap5firstorderamplitude}
    \dot{a}_j&=-\frac{\gamma_j}{2}a_j-\langle H^{\text{fb},0}_j\sin{(\omega_jt+\chi_j)\rangle_T}\nonumber\\
    &~~~~-\langle\xi_j\sin{(\omega_jt+\chi_j)}\rangle_T,\\
    \dot{\chi}_j&=-\frac{1}{a_j}\langle H^{\text{fb},0}_j\cos{(\omega_jt+\chi_j)}\rangle_T\nonumber\\
    &~~~~-\langle\frac{\xi_j}{a_j}\cos{(\omega_jt+\chi_j)}\rangle_T.
\end{align}
To compute the time-averaged feedback force, we first express $F_I$ and $F_D$ as:
\begin{align}
    F_I&=\sum_j\frac{a_j}{\omega_j}\left[\sin{(\omega_j t+\chi_j)}-\sin{\chi_j}\right]\nonumber\\
    &\approx\sum_j\frac{a_j}{\omega_j}\sin{(\omega_j t+\chi_j)},\\
    F_D&=-\sum_ja_j\omega_j\sin{(\omega_jt+\chi_j)},
    \label{eq:fi}
\end{align}
where the second term of $F_I$ is omitted, assuming $\sum_j\frac{a_j}{\omega_j}\sin{\chi_j}\approx 0$ due to the uncorrelated phases of the mechanical modes. We are interested in the amplitude equations. Plugging $F_I$ and $F_D$ into the time average of $\langle H^{\text{fb},0}_j\sin{(\omega_jt+\chi_j)}\rangle_T$, we obtain the nonzero terms:
\begin{itemize}
    \item For the first term $F_I^2F_D$. The nonzero terms are
    \begin{align}
        -\sum_{i\neq j}\frac{a_i^2}{\omega_i^2}a_j\omega_j\langle\sin^2{(\omega_it+\chi_i)}\sin^2{(\omega_jt+\chi_j)}\rangle_T\nonumber\\
        =-\frac{1}{4}\sum_{i\neq j}\frac{a_i^2}{\omega_i^2}a_j\omega_j,\\
        -2\sum_{i\neq j}\frac{a_i^2}{\omega_j}a_j\langle\sin^2{(\omega_it+\chi_i)}\sin^2{(\omega_jt+\chi_j)}\rangle_T\nonumber\\
        =-\frac{1}{2}\sum_{i\neq j}\frac{a_i^2}{\omega_j}a_j,
    \end{align}
    and
    \begin{equation}
        -\frac{a_j^3}{\omega_j}\langle\sin^4{(\omega_jt+\chi_j)}\rangle_T=-\frac{3}{8}\frac{a_j^3}{\omega_j}.
    \end{equation}
    
    \item For the second term $3Q^2F_I$, the nonzero terms are
    \begin{align}
        3\sum_{i\neq j}a_i^2\frac{a_j}{\omega_j}\langle\cos^2{(\omega_it+\chi_i)}\sin^2{(\omega_jt+\chi_j)}\rangle_T\nonumber\\
        =\frac{3}{4}\sum_{i\neq j}a_i^2\frac{a_j}{\omega_j},
    \end{align}
    and
    \begin{equation}
        3\frac{a_j^3}{\omega_j}\langle\cos^2{(\omega_jt+\chi_j)}\sin^2{(\omega_jt+\chi_j)}\rangle_T=\frac{3}{8}\frac{a_j^3}{\omega_j}
    \end{equation}
\end{itemize}
In this derivation, we have assumed incommensurate eigenfrequencies. Hence, the time average takes the form
\begin{equation}
    \langle H^{\text{fb}}_j\sin{(\omega_jt+\chi_j)}\rangle_T=-\frac{g_j\omega_{\text{fb}}}{4}\sum_{i\neq j}\frac{\omega_i^2-\omega_j^2}{\omega_i^2\omega_j}a_i^2a_j.
    \label{eq:feedbackaverage}
\end{equation}
Plugging Eq. (\ref{eq:feedbackaverage}) into Eq. (\ref{eq:chap5firstorderamplitude}), we obtain the amplitude equations (\ref{eq:amplitudeequations}) in the main text.

\section{Oscillator under a simple drive}
\label{appendix:simpledrive}
To compare the feedback-induced phonon condensation with the most direct alternative protocol, we consider a simple harmonic drive of the form $H_{\text{drive}}^{(j)}=f_j\cos{(\omega_d t)}$. The equations of motion are
\begin{align}
    \dot{Q}_j&=\omega_jP_j,\nonumber\\
    \dot{P}_j&=-\omega_jQ_j-\gamma_jP_j+\xi_j+f_j\cos{(\omega_d t)}.
\end{align}
Because the drive is linear, it does not induce interactions between different modes. The solution can therefore be decomposed into a deterministic part set by the drive and a stochastic part generated by thermal noise:
\begin{align}
    Q_j(t)&=Q_{j,d}(t)+Q_{j,\text{th}}(t),\\
    P_j(t)&=P_{j,d}(t)+P_{j,\text{th}}(t),
\end{align}
where $Q_{j,d}(t)$ and $P_{j,d}(t)$ denote the deterministic responses to the drive, while $Q_{j,\text{th}}(t)$ and $P_{j,\text{th}}(t)$ denote the stochastic thermal contributions. The deterministic response is
\begin{align}
    Q_{j,d}(t)&=\frac{f_j\omega_j}{\sqrt{(\omega_j^2-\omega_d^2)^2+\gamma_j^2\omega_d^2}}\cos{(\omega_d t+\phi_j)},\\
    P_{j,d}(t)&=-\frac{f_j\omega_d}{\sqrt{(\omega_j^2-\omega_d^2)^2+\gamma_j^2\omega_d^2}}\sin{(\omega_d t+\phi_j)},
\end{align}
where $\phi_j$ is the phase shift. The stochastic part still satisfies the equipartition theorem, namely $\langle Q_{j,\text{th}}^2\rangle=\langle P_{j,\text{th}}^2\rangle=1$. However, this does \emph{not} mean that the energy variance is unaffected by the drive, because the energy is a quadratic function of the total displacement and momentum. Once the deterministic and thermal parts are combined, cross terms between the coherent response and the thermal fluctuations contribute to the second moment of the energy. Using standard Gaussian moments, we have $\langle Q_{j,\text{th}}^4\rangle=3\langle Q_{j,\text{th}}^2\rangle^2=3$, while all odd moments vanish. The total energy of mode $j$ is therefore
\begin{align}
    \langle E_j\rangle &= \frac{1}{2}\langle Q_j^2 + P_j^2\rangle \nonumber\\
    &= \frac{1}{2}\left(Q_{j,d}^2 + P_{j,d}^2 \right)+ \frac{1}{2}\langle Q_{j,\text{th}}^2 + P_{j,\text{th}}^2\rangle \nonumber\\
    & = E_{j,d} + 1.
\end{align}
The square of the energy is given by
\begin{align}
    \langle E_j^2\rangle &= \frac{1}{4}\langle (Q_j^2 + P_j^2)^2 \rangle \nonumber\\
    & = E_{j,d}^2 + 4E_{j,d} + 2.
\end{align}
Hence, the variance of the energy is given by
\begin{align}
    \text{Var}(E_j) &= \langle E_j^2\rangle - \langle E_j\rangle^2 \nonumber\\
    & = 2E_{j,d} + 1.
\end{align}
These expressions show that, under a simple resonant drive, the mean energy and the energy variance increase together. The corresponding $g^{(2)}(0)$ function is
\begin{align}
    g^{(2)}(0) = \frac{\langle E_j^2\rangle}{\langle E_j\rangle^2} = 1 + \frac{2E_{j,d} + 1}{(E_{j,d} + 1)^2}.
\end{align}
In the strong-drive limit, $E_{j,d}\gg 1$, we have $g^{(2)}(0) \approx 1 + \frac{2}{E_{j,d}}$, so $g^{(2)}(0)$ approaches 1 from above as the drive strength increases.

The energy distribution of the driven mode can also be obtained analytically. In steady state, the phase-space distribution is a Gaussian centered on the deterministic response,
\begin{align}
    P(Q_j,P_j,t) = \frac{1}{2\pi}e^{-\frac{1}{2}[(Q_j-Q_{j,d})^2+(P_j-P_{j,d})^2]}.
\end{align}
Transforming to polar coordinates yields the energy distribution, which follows a noncentral $\chi^2$ distribution,
\begin{align}
    P(E_j,t) = e^{-(E_j + E_{j,d})}I_0\left(2\sqrt{E_j E_{j,d}}\right),
\end{align}
where $I_0$ is the modified Bessel function of the first kind.

\section{Numerical simulation of spectral density}
\label{appendix:spectraldensity}
 The noise power spectral density $S_{Q_1Q_1}(\omega)$ is the Fourier transform of the autocorrelation function (\ref{eq:sqq}). To numerically calculate $S_{Q_1Q_1}(\omega)$, we employ an efficient method rather than directly using the definition. For each trajectory, we numerically integrate Eq. (\ref{eq:eoms}) from $t=0$ to $t=T$ to obtain the displacement $Q_1(t)$. We then calculate the Fourier transform of the displacement
\begin{align}
    \tilde{Q}_1(\omega)=\frac{1}{\sqrt{T}}\int_0^{T}dtQ_1(t)e^{i\omega t}
\end{align}
In the large $T$ limit, the spectral density $S_{Q_1Q_1}(\omega)$ is equal to the ensemble average of $|\tilde{Q}_1(\omega)|^2$, i.e., 
\begin{align}
    S_{Q_1Q_1}(\omega)=\text{lim}_{T\to\infty}\langle|\tilde{Q}_1(\omega)|^2\rangle
\end{align}

\section{Comparison of Fr\"{o}hlich's model and our model}
\label{appendix:comparison}
The rate equations of phonon numbers in Fr\"{o}hlich's model are given by
\begin{align}
    \dot{n}_j&=s-\gamma_j(n_j-\bar{n}_{j,\text{th}})\nonumber\\
    &+\chi\sum_i[(n_j+1)n_i-n_j(1+n_i)e^{\hbar(\omega_j-\omega_i)/k_BT}], \label{eq:frohlichoriginal}
\end{align}
where $s$ is the external pumping, $\chi$ is the coupling strength of two-phonon process, $\bar{n}_{j,\text{th}}$ is the thermal phonon number. In the limit of large phonon number $n_j\gg1$, the equations are simplified as
\begin{align}
    \dot{n}_j=&s-\gamma_j(n_j-\bar{n}_{j,\text{th}})+\chi\sum_i[1-e^{\hbar(\omega_j-\omega_i)/k_BT}]n_in_j. \label{eq:frohlichrate}
\end{align}
Recently, a proposal was made to realize Fr\"{o}hlich condensation in optomechanical systems \cite{zheng2021froh}. The modified rate equations of phonon numbers are given by
\begin{align}
    \dot{n}_j=&-\gamma_j(n_j-\bar{n}_{j,\text{th}})\nonumber\\
    &+\sum_{i\neq j}4U_{i,j}^2\left[\Gamma(\omega_i-\omega_j)-\Gamma(\omega_j-\omega_i)\right]n_in_j,
    \label{eq:optorate}
\end{align}
where $U_{i,j}$ is a coefficient and $\Gamma(\omega)$ is a function of frequency. To compare the amplitude equations (\ref{eq:amplitudeequations}) with Eq. (\ref{eq:frohlichrate}) and (\ref{eq:optorate}), we need to convert Eq. (\ref{eq:amplitudeequations}) to the rate equations of $\langle |a_j(t)|^2\rangle$. The formal solution of Eq. (\ref{eq:amplitudeequations}) is given by
\begin{align}
    &a_j(t)\nonumber\\
    &=\int_{-\infty}^{t}dse^{-\frac{\gamma_j}{2}(t-s)+\sum_i\frac{g_j\omega_{\text{fb}}}{4\omega_i^2\omega_j}(\omega_i^2-\omega_j^2)\int_{s}^tdt^{\prime}|a_i(t^{\prime})|^2}\Xi_j(s)
    \label{eq:formalaj}
\end{align}
From Eq. (\ref{eq:formalaj}) we can get the formal solution of $\langle |a_j(t)|^2\rangle$,
\begin{align}
    &\langle |a_j(t)|^2\rangle\nonumber\\
    &\approx2\gamma_j\int_{-\infty}^{t}ds\langle e^{-\gamma_j(t-s)+\sum_i\frac{g_j\omega_{\text{fb}}}{2\omega_i^2\omega_j}(\omega_i^2-\omega_j^2)\int_{s}^tdt^{\prime}|a_i(t^{\prime})|^2}\rangle,
    \label{eq:formalaj2}
\end{align}
where we have used Eq. (\ref{eq:correlation}) and decorrelation approximation. Hence, the rate equations of $\langle |a_j|^2\rangle$ are given by
\begin{align}
    \frac{d\langle |a_j|^2\rangle}{dt}\approx &-\gamma_j(\langle|a_j|^2\rangle-2)\nonumber\\
    &+\sum_i\frac{g_j\omega_{\text{fb}}}{2\omega_i^2\omega_j}(\omega_i^2-\omega_j^2)\langle|a_i|^2\rangle\langle|a_j|^2\rangle
    \label{eq:aj2equations}
\end{align}
under decorrelation approximation. In the high temperature limit, the phonon numbers are determined by $n_j=\langle |a_j|^2\rangle k_BT/(2\hbar\omega_j)$. From Eq. (\ref{eq:aj2equations}), the rate equations of phonon numbers in our model are then given by
\begin{align}
    \dot{n}_j\approx&-\gamma_j(n_j-\bar{n}_{j,\text{th}})+\sum_i\frac{\hbar g_j\omega_{\text{fb}}}{k_BT\omega_i\omega_j}(\omega_i^2-\omega_j^2)n_i n_j
    \label{eq:rateequationourmodel}
\end{align}
Equation (\ref{eq:rateequationourmodel}) shares the same form as Eqs. (\ref{eq:frohlichrate}) and (\ref{eq:optorate}), differing only in the coupling coefficient for the $n_in_j$ term and the absence of external pumping compared to Eq. (\ref{eq:frohlichoriginal}).

\bibliography{reference}

@article{wen2020coherent,
  title={A coherent nanomechanical oscillator driven by single-electron tunnelling},
  author={Wen, Yutian and Ares, N and Schupp, FJ and Pei, T and Briggs, GAD and Laird, EA},
  journal={Nat. Physics},
  volume={16},
  pages={75--82},
  year={2020},
  doi={10.1038/s41567-019-0683-5},
  publisher={Nature Publishing Group}
}

@article{pettit2019optical,
  title={An optical tweezer phonon laser},
  author={Pettit, Robert M and Ge, Wenchao and Kumar, P and Luntz-Martin, Danika R and Schultz, Justin T and Neukirch, Levi P and Bhattacharya, Mishkat and Vamivakas, A Nick},
  journal={Nat. Photonics},
  volume={13},
  number={6},
  pages={402--405},
  year={2019},
  doi={10.1038/s41566-019-0395-5},
  publisher={Nature Publishing Group}
}

@article{ohta2017feedback,
  title={Feedback control of multiple mechanical modes in coupled micromechanical resonators},
  author={Ohta, Ryuichi and Okamoto, Hajime and Yamaguchi, Hiroshi},
  journal={Appl. Phys. Lett.},
  volume={110},
  number={5},
  pages={053106},
  year={2017},
  doi={10.1063/1.4975207},
  publisher={AIP Publishing LLC}
}

@article{rugar2004single,
  title={Single spin detection by magnetic resonance force microscopy},
  author={Rugar, Daniel and Budakian, Raffi and Mamin, HJ and Chui, BW},
  journal={Nature},
  volume={430},
  number={6997},
  pages={329--332},
  year={2004},
  doi={10.1038/nature02658},
  publisher={Nature Publishing Group}
}

@article{yang2006zeptogram,
  title={Zeptogram-scale nanomechanical mass sensing},
  author={Yang, Ya-Tang and Callegari, Carlo and Feng, XL and Ekinci, Kamil L and Roukes, Michael L},
  journal={Nano Lett.},
  volume={6},
  number={4},
  pages={583--586},
  year={2006},
  doi={10.1021/nl052134m},
  publisher={ACS Publications}
}

@article{burg2007weighing,
  title={Weighing of biomolecules, single cells and single nanoparticles in fluid},
  author={Burg, Thomas P and Godin, Michel and Knudsen, Scott M and Shen, Wenjiang and Carlson, Greg and Foster, John S and Babcock, Ken and Manalis, Scott R},
  journal={Nature},
  volume={446},
  number={7139},
  pages={1066--1069},
  year={2007},
  doi={10.1038/nature05741},
  publisher={Nature Publishing Group}
}

@article{masmanidis2007multifunctional,
  title={Multifunctional nanomechanical systems via tunably coupled piezoelectric actuation},
  author={Masmanidis, Sotiris C and Karabalin, Rassul B and De Vlaminck, Iwijn and Borghs, Gustaaf and Freeman, Mark R and Roukes, Michael L},
  journal={Science},
  volume={317},
  number={5839},
  pages={780--783},
  year={2007},
  doi={10.1126/science.1144793},
  publisher={American Association for the Advancement of Science}
}

@article{feng2008self,
  title={A self-sustaining ultrahigh-frequency nanoelectromechanical oscillator},
  author={Feng, XL and White, CJ and Hajimiri, A and Roukes, Michael L},
  journal={Nat. Nanotechnology},
  volume={3},
  number={6},
  pages={342--346},
  year={2008},
  doi={10.1038/nnano.2008.125},
  publisher={Nature Publishing Group}
}

@article{tadokoro2018driven,
  title={Driven nonlinear nanomechanical resonators as digital signal detectors},
  author={Tadokoro, Yukihiro and Tanaka, Hiroya and Dykman, MI},
  journal={Sci. Rep.},
  volume={8},
  pages={11284},
  year={2018},
  doi={10.1038/s41598-018-29572-7},
  publisher={Nature Publishing Group}
}

@article{mahboob2008bit,
  title={Bit storage and bit flip operations in an electromechanical oscillator},
  author={Mahboob, I and Yamaguchi, H},
  journal={Nat. Nanotechnology},
  volume={3},
  number={5},
  pages={275--279},
  year={2008},
  doi={10.1038/nnano.2008.84},
  publisher={Nature Publishing Group}
}

@article{unterreithmeier2009universal,
  title={Universal transduction scheme for nanomechanical systems based on dielectric forces},
  author={Unterreithmeier, Quirin P and Weig, Eva M and Kotthaus, J{\"o}rg P},
  journal={Nature},
  volume={458},
  number={7241},
  pages={1001--1004},
  year={2009},
  doi={10.1038/nature07932},
  publisher={Nature Publishing Group}
}

@article{shekhawat2005nanoscale,
  title={Nanoscale imaging of buried structures via scanning near-field ultrasound holography},
  author={Shekhawat, Gajendra S and Dravid, Vinayak P},
  journal={Science},
  volume={310},
  number={5745},
  pages={89--92},
  year={2005},
  doi={10.1126/science.1117694},
  publisher={American Association for the Advancement of Science}
}

@article{tetard2008imaging,
  title={Imaging nanoparticles in cells by nanomechanical holography},
  author={Tetard, Laurene and Passian, Ali and Venmar, Katherine T and Lynch, Rachel M and Voy, Brynn H and Shekhawat, Gajendra and Dravid, Vinayak P and Thundat, Thomas},
  journal={Nat. Nanotechnology},
  volume={3},
  number={8},
  pages={501--505},
  year={2008},
  doi={10.1038/nnano.2008.162},
  publisher={Nature Publishing Group}
}

@article{tamayo2001high,
  title={High-{Q} dynamic force microscopy in liquid and its application to living cells},
  author={Tamayo, J and Humphris, ADL and Owen, RJ and Miles, MJ},
  journal={Biophys. J.},
  volume={81},
  pages={526--537},
  year={2001},
  doi={10.1016/S0006-3495(01)75719-0},
  publisher={Elsevier}
}

@article{frohlich1968long,
  title={Long-range coherence and energy storage in biological systems},
  author={{F}r{\"o}hlich, Herbert},
  journal={Int. J. Quantum Chem.},
  volume={2},
  number={5},
  pages={641--649},
  year={1968},
  doi={10.1002/qua.560020505},
  publisher={Wiley Online Library}
}

@article{frohlich1968bose,
  title={Bose condensation of strongly excited longitudinal electric modes},
  author={{F}r{\"o}hlich, Herbert},
  journal={Phys. Lett. A},
  volume={26},
  number={9},
  pages={402--403},
  year={1968},
  doi={10.1016/0375-9601(68)90242-9},
  publisher={Elsevier}
}

@article{frohlich1970long,
  title={Long range coherence and the action of enzymes},
  author={{F}r{\"o}hlich, H},
  journal={Nature},
  volume={228},
  number={5276},
  pages={1093--1093},
  year={1970},
  doi={10.1038/2281093a0},
  publisher={Nature Publishing Group}
}

@article{reimers2009weak,
  title={Weak, strong, and coherent regimes of {F}r{\"o}hlich condensation and their applications to terahertz medicine and quantum consciousness},
  author={Reimers, Jeffrey R and McKemmish, Laura K and McKenzie, Ross H and Mark, Alan E and Hush, Noel S},
  journal={Proc. Natl. Acad. Sci. U.S.A.},
  volume={106},
  number={11},
  pages={4219--4224},
  year={2009},
  doi={10.1073/pnas.0806273106},
  publisher={National Academy of Sciences}
}

@article{zhang2019quantum,
  title={Quantum Fluctuations in the {F}r{\"o}hlich Condensate of Molecular Vibrations Driven Far From Equilibrium},
  author={Zhang, Zhedong and Agarwal, Girish S and Scully, Marlan O},
  journal={Phys. Rev. Lett.},
  volume={122},
  number={15},
  pages={158101},
  year={2019},
  doi={10.1103/PhysRevLett.122.158101},
  publisher={APS}
}

@book{pikovsky2003synchronization,
  title={Synchronization: A Universal Concept in Nonlinear Sciences},
  author={Pikovsky, Arkady and Rosenblum, Michael and Kurths, J{\"u}rgen},
  series={Cambridge Nonlinear Science Series},
  volume={12},
  year={2001},
  doi={10.1017/CBO9780511755743},
  publisher={Cambridge University Press}
}

@article{zhang2018mode,
  title={Mode competition and hopping in optomechanical nano-oscillators},
  author={Zhang, Xingwang and Lin, Tong and Tian, Feng and Du, Han and Zou, Yongchao and Chau, Fook Siong and Zhou, Guangya},
  journal={Appl. Phys. Lett.},
  volume={112},
  number={15},
  pages={153502},
  year={2018},
  doi={10.1063/1.5008664},
  publisher={AIP Publishing LLC}
}

@article{kemiktarak2014mode,
  title={Mode competition and anomalous cooling in a multimode phonon laser},
  author={Kemiktarak, Utku and Durand, Mathieu and Metcalfe, Michael and Lawall, John},
  journal={Phys. Rev. Lett.},
  volume={113},
  number={3},
  pages={030802},
  year={2014},
  doi={10.1103/PhysRevLett.113.030802},
  publisher={APS}
}

@article{jonsson2008self,
  title={Self-organization of irregular nanoelectromechanical vibrations in multimode shuttle structures},
  author={Jonsson, L Magnus and Santandrea, Fabio and Gorelik, Leonid Y and Shekhter, Robert I and Jonson, Mats},
  journal={Phys. Rev. Lett.},
  volume={100},
  number={18},
  pages={186802},
  year={2008},
  doi={10.1103/PhysRevLett.100.186802},
  publisher={APS}
}

@article{sommer2019partial,
  title={Partial optomechanical refrigeration via multimode cold-damping feedback},
  author={Sommer, Christian and Genes, Claudiu},
  journal={Phys. Rev. Lett.},
  volume={123},
  number={20},
  pages={203605},
  year={2019},
  doi={10.1103/PhysRevLett.123.203605},
  publisher={APS}
}

@article{preto2017semi,
  title={Semi-classical statistical description of {F}r{\"o}hlich condensation},
  author={Preto, Jordane},
  journal={J. Biol. Phys.},
  volume={43},
  number={2},
  pages={167--184},
  year={2017},
  doi={10.1007/s10867-017-9442-y},
  publisher={Springer}
}

@article{zheng2021froh,
  title = {Fr\"ohlich condensate of phonons in optomechanical systems},
  author = {Zheng, Xu and Li, Baowen},
  journal = {Phys. Rev. A},
  volume = {104},
  issue = {4},
  pages = {043512},
  numpages = {12},
  year = {2021},
  month = {Oct},
  doi = {10.1103/PhysRevA.104.043512},
  publisher = {American Physical Society}

}

@article{sommer2020prospects,
  title={Prospects of reinforcement learning for the simultaneous damping of many mechanical modes},
  author={Sommer, Christian and Asjad, Muhammad and Genes, Claudiu},
  journal={Sci. Rep.},
  volume={10},
  pages={2623},
  year={2020},
  doi={10.1038/s41598-020-59435-z},
  publisher={Nature Publishing Group}
}

@article{wu1977bose,
  title={Bose condensation in biosystems},
  author={Wu, TM and Austin, Steven},
  journal={Phys. Lett. A},
  volume={64},
  pages={151--152},
  year={1977},
  doi={10.1016/0375-9601(77)90560-6},
  publisher={Elsevier}
}

@article{wu1978cooperative,
  title={Cooperative behavior in biological systems},
  author={Wu, TM and Austin, Steven},
  journal={Phys. Lett. A},
  volume={65},
  pages={74--76},
  year={1978},
  doi={10.1016/0375-9601(78)90137-8},
  publisher={Elsevier}
}

@article{wu1981frohlich,
  title={{F}r{\"o}hlich's model of {B}ose condensation in biological systems},
  author={Wu, TM and Austin, Steven J},
  journal={J. Biol. Phys.},
  volume={9},
  number={2},
  pages={97--107},
  year={1981},
  doi={10.1007/BF01987286},
  publisher={Springer}
}

@article{dekorsy2000coherent,
  title={Coherent phonons in condensed media},
  author={Dekorsy, Thomas and Cho, Gyu Cheon and Kurz, Heinrich},
  journal={Light scattering in solids VIII},
  volume={76},
  pages={169--209},
  year={2000},
  doi={10.1007/bfb0084242},
  publisher={Springer}
}

@article{ruello2015physical,
  title={Physical mechanisms of coherent acoustic phonons generation by ultrafast laser action},
  author={Ruello, Pascal and Gusev, Vitalyi E},
  journal={Ultrasonics},
  volume={56},
  pages={21--35},
  year={2015},
  doi={10.1016/j.ultras.2014.06.004},
  publisher={Elsevier}
}

@article{poyser2015coherent,
  title={Coherent phonon optics in a chip with an electrically controlled active device},
  author={Poyser, Caroline L and Akimov, Andrey V and Campion, Richard P and Kent, Anthony J},
  journal={Sci. Rep.},
  volume={5},
  pages={8279},
  year={2015},
  doi={10.1038/srep08279},
  publisher={Nature Publishing Group}
}

@inproceedings{ruskov2012coherent,
  title={Coherent phonons as a new element of quantum computing and devices},
  author={Ruskov, Rusko and Tahan, Charles},
  booktitle={J. Phys. Conf. Ser.},
  volume={398},
  pages={012011},
  year={2012},
  doi={10.1088/1742-6596/398/1/012011},
  organization={IOP Publishing}
}

@article{gustafsson2014propagating,
  title={Propagating phonons coupled to an artificial atom},
  author={Gustafsson, Martin V and Aref, Thomas and Kockum, Anton Frisk and Ekstr{\"o}m, Maria K and Johansson, G{\"o}ran and Delsing, Per},
  journal={Science},
  volume={346},
  number={6206},
  pages={207--211},
  year={2014},
  doi={10.1126/science.1257219},
  publisher={American Association for the Advancement of Science}
}

@article{bienfait2019phonon,
  title={Phonon-mediated quantum state transfer and remote qubit entanglement},
  author={Bienfait, Audrey and Satzinger, Kevin J and Zhong, YP and Chang, H-S and Chou, M-H and Conner, Chris R and Dumur, {\'E} and Grebel, Joel and Peairs, Gregory A and Povey, Rhys G and others},
  journal={Science},
  volume={364},
  number={6438},
  pages={368--371},
  year={2019},
  doi={10.1126/science.aaw8415},
  publisher={American Association for the Advancement of Science}
}

@article{liu2020review,
  title={A review of acoustic metamaterials and phononic crystals},
  author={Liu, Junyi and Guo, Hanbei and Wang, Ting},
  journal={Crystals},
  volume={10},
  number={4},
  pages={305},
  year={2020},
  doi={10.3390/cryst10040305},
  publisher={Multidisciplinary Digital Publishing Institute}
}

@article{takabatake2014phonon,
  title={Phonon-glass electron-crystal thermoelectric clathrates: Experiments and theory},
  author={Takabatake, Toshiro and Suekuni, Koichiro and Nakayama, Tsuneyoshi and Kaneshita, Eiji},
  journal={Rev. Mod. Phys.},
  volume={86},
  number={2},
  pages={669},
  year={2014},
  doi={10.1103/RevModPhys.86.669},
  publisher={APS}
}

@article{li2012colloquium,
  title={Colloquium: Phononics: Manipulating heat flow with electronic analogs and beyond},
  author={Li, Nianbei and Ren, Jie and Wang, Lei and Zhang, Gang and H{\"a}nggi, Peter and Li, Baowen},
  journal={Rev. Mod. Phys.},
  volume={84},
  number={3},
  pages={1045},
  year={2012},
  doi={10.1103/RevModPhys.84.1045},
  publisher={APS}
}

@article{li2021transforming,
  title={Transforming heat transfer with thermal metamaterials and devices},
  author={Li, Ying and Li, Wei and Han, Tiancheng and Zheng, Xu and Li, Jiaxin and Li, Baowen and Fan, Shanhui and Qiu, Cheng-Wei},
  journal={Nat. Rev. Mater.},
  volume={6},
  number={6},
  pages={488--507},
  year={2021},
  doi={10.1038/s41578-021-00283-2},
  publisher={Nature Publishing Group}
}

@book{landau1986course,
  title={Theory of Elasticity},
  author={Landau, L. D. and Lifshitz, E. M.},
  series={Course of Theoretical Physics},
  volume={7},
  edition={3rd},
  year={1986},
  url={https://shop.elsevier.com/books/theory-of-elasticity/landau/978-0-08-057069-3},
  publisher={Pergamon Press}
}

@article{rackauckas2017differentialequations,
  title={{DifferentialEquations.jl} -- A performant and feature-rich ecosystem for solving differential equations in Julia},
  author={Rackauckas, Christopher and Nie, Qing},
  journal={J. Open Res. Softw.},
  volume={5},
  pages={15},
  year={2017},
  doi={10.5334/jors.151},
  publisher={Ubiquity Press, Ltd.}
}

@article{hu1996squeezed,
  title={Squeezed Phonon States: Modulating Quantum Fluctuations of Atomic Displacements},
  author={Hu, X and Nori, Franco},
  journal={Phys. Rev. Lett.},
  volume={76},
  number={13},
  pages={2294},
  year={1996},
  doi={10.1103/PhysRevLett.76.2294},
  publisher={APS}
}

@article{poggio2007feedback,
  title = {Feedback Cooling of a Cantilever's Fundamental Mode below 5 mK},
  author = {Poggio, M. and Degen, C. L. and Mamin, H. J. and Rugar, D.},
  journal = {Phys. Rev. Lett.},
  volume = {99},
  issue = {1},
  pages = {017201},
  numpages = {4},
  year = {2007},
  month = {Jul},
  doi = {10.1103/PhysRevLett.99.017201},
  publisher = {American Physical Society}
}

@article{wang2023fastfeedback,
  title={Fast Feedback Control of Mechanical Motion Using Circuit Optomechanics},
  author = {Wang, Cheng and Banniard, Louise and de L\'epinay, Laure Mercier and Sillanp\"a\"a, Mika A.},
  journal = {Phys. Rev. Appl.},
  volume = {19},
  issue = {5},
  pages = {054091},
  numpages = {14},
  year = {2023},
  month = {May},
  doi = {10.1103/PhysRevApplied.19.054091},
  publisher = {American Physical Society}
}

@article{Guo2019feedback,
  title = {Feedback Cooling of a Room Temperature Mechanical Oscillator close to its Motional Ground State},
  author = {Guo, Jingkun and Norte, Richard and Gr\"oblacher, Simon},
  journal = {Phys. Rev. Lett.},
  volume = {123},
  issue = {22},
  pages = {223602},
  numpages = {6},
  year = {2019},
  month = {Nov},
  doi = {10.1103/PhysRevLett.123.223602},
  publisher = {American Physical Society},
}

@article{guo2023active,
  title={Active-feedback quantum control of an integrated low-frequency mechanical resonator},
  author={Guo, Jingkun and Chang, Jin and Yao, Xiong and Gr{\"o}blacher, Simon},
  journal={Nature Communications},
  volume={14},
  pages={4721},
  year={2023},
  doi={10.1038/s41467-023-40442-3},
  publisher={Nature Publishing Group UK London}
}

@article{dong2019more,
  title={More than 20 W fiber-based continuous-wave single frequency laser at 780 nm},
  author={Dong, Jinyan and Zeng, Xin and Cui, Shuzhen and Zhou, Jiaqi and Feng, Yan},
  journal={Optics Express},
  volume={27},
  number={24},
  pages={35362--35367},
  year={2019},
  doi={10.1364/OE.27.035362},
  publisher={Optical Society of America}
}

\end{document}